\documentclass[%
 aip,
 jmp,%
 amsmath,amssymb,
preprint,%
]{revtex4-1}

\usepackage{graphicx}
\usepackage{dcolumn}
\usepackage{bm}

\begin{document}

\preprint{AIP/123-QED}

\title[]{A capacitive displacement system for studying  the  piezoelectric strain and its temperature variation}

\author{Desheng Fu$^{1,2,3}$ and Eiki Kakihara$^{2}$\\
$^1$Department of Electronics and Materials Science, Faculty of
Engineering, Shizuoka University, 3-5-1 Johoku, Naka-ku, Hamamatsu
432-8561, Japan.\\
$^2$Department of Engineering, Graduate School of Integrated Science
and Technology, Shizuoka University, 3-5-1 Johoku, Hamamatsu
432-8561,
Japan.\\
$^3$Department of Optoelectronics and Nanostructure Science,
Graduate School of Science and Technology, 3-5-1 Johoku, Naka-ku,
Hamamatsu 432-8011, Japan.}


\email{fu.tokusho@shizuoka.ac.jp}
\homepage{https://wwp.shizuoka.ac.jp/desheng-fu/}



\date{\today}

\begin{abstract}
A capacitive displacement system was constructed to measure the electric-field-induced piezoelectric strain  in the simple  form of   either  a  bulk or  thin film.  The system can determine an  AC displacement of   2 pm precisely  by using a lock-in
detection, and  can measure the large  displacement within a range of $\pm$ 25 $\mu$m with a sub-nanometer resolution. The system  can also be used to measure the variation in strain  within a  temperature  range of  210 - 450 K, allowing  the  evaluation of the  temperature coefficient of  a piezoelectric constant  and the studies on  the effects of a phase transition  on the piezoelectric response. Experimental results on quartz, PZT ceramics and  thin films, and BaTiO$_3$ confirm the capabilities of the developed system.
\end{abstract}

\pacs{77.65.-j, 77.84.Dy, 77.55.+f, 77.80.Bh, 07.10.Pz, 81.70.-q}
\maketitle

 \section{Introduction}

Piezoelectrics convert  mechanical energy into electric energy or vise versa, and show direct or converse
effects.\cite{Curie,Jaffe,Lines,Uchino} In a direct effect, electric charges are produced at the surface of the piezoelectric crystal
in proportion to the applied force. Conversely, as a suitably oriented electric field is applied, the crystal changes shape (strains) in proportion to the electric field. Such properties provide piezoelectrics with a wide spectra of applications such as medical ultrasound imaging, ultra-precise positioning,  and sonar sensing. With a combination of modern integration- and nano-technologies, piezoelectrics  can also be used as nano-sensors, nano-actuators,  or nano-transducers in modern smart devices.\cite{Muralt1995,Muralt1997,Damjanovic,DeVoe}  Therefore, there has been increasing interest in developing novel applications of piezoelectric materials such as  current piezoelectric
energy harvesting,  or interest in exploring green materials to replace  Pb-containing  piezoelectric materials.\cite{ZhangS1, McKinstry, Liu, Acosta, FuAPL2007, FuPRL2008,FuAPL2008,Rodel, FuJPC2011,FuAPL2011,FuJPC2013}

Various techniques have been developed  to study the piezoelectric properties.\cite{IEEE,Berlincourt,Bottom,Pan,Lefki,Kholkin,Kanno,Christman,Xu,Durkan,Kuffer,FuJJAP,Park}
The resonance technique is the most widely used method\cite{IEEE} with a measurement accuracy of approximately  $\pm$ 2\%.\cite{Fialka}  However, it has special requirements for the shape and dimensions of the sample for different   vibrational modes, and thus it is difficult to be used for   thin-film measurements owning  to  shape and dimension limitations. The  Berlincourt meter or $d_{33}$-meter \cite{Berlincourt}  based on a  direct effect are also widely used to measure the $d_{33}$ value of the bulk, which also shows a high measurement accuracy of approximately  $\pm$ 1\%.\cite{Stewart}  With this method, an AC force and a pre-stress are applied to the sample\cite{Stewart}, which may result in  damage to a  thin film. Thus, it is  unsuitable for thin film measurements. Another commonly used technique is the  interferometry method, which is  based on the converse effect for determining  the sample displacement  driven by an AC field. The  interferometry has a pm-resolution  to  precisely determine the small AC displacement and can be used for both bulk and thin films.\cite{Bottom,Pan, Kholkin} However, the  interferometry  method is difficult to  measure  a  large displacement in the micro-meter scale.

Herein, we demonstrate a capacitive displacement method that  can be used  to reliably and simply measure the electric-field-induced strain in  the form of either a bulk or  film. We also show that our system can detect an AC displacement of as small as 2 pm, which is comparable with the resolution of the interferometry method.\cite{Pan} It  can also measure a large displacement within a range of $\pm$ 25 $\mu$m with a sub-nanometer resolution. Moreover, the developed system can be used to measure the variation of strain with temperature in a wide temperature range of  210  to 450 K, allowing an estimation of  the temperature dependence of the strain or  studies on the effects of the  phase transition on the piezoelectric properties.

 \section{System Design}
\begin{figure}
\includegraphics[width=12cm]{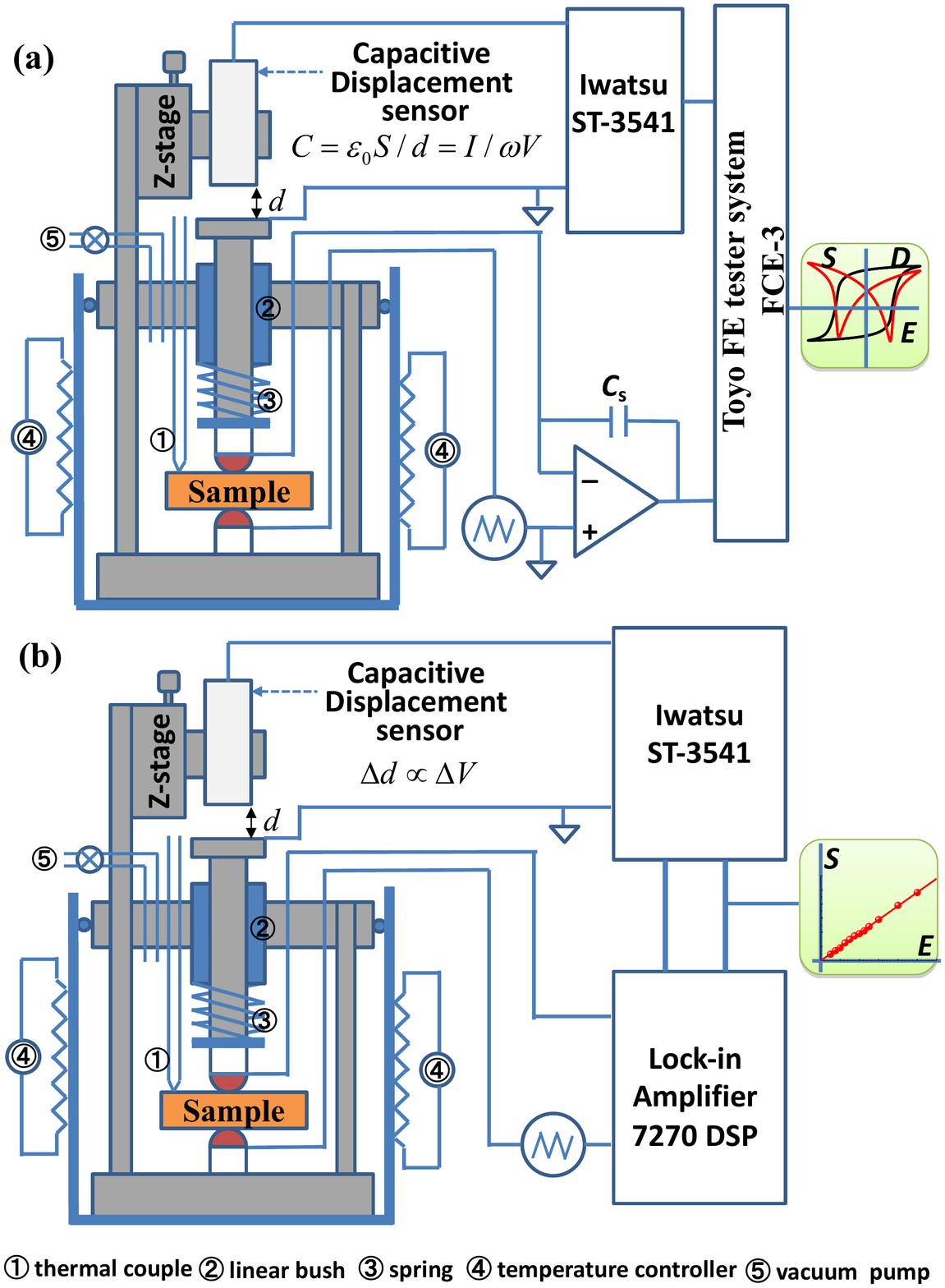}
\caption{\label{Fig1} A block diagram of the system. (a) System operating in a normal mode, allowing the simultaneous measurements  the  dielectric and displacement hysteresis loops for a bipolar electric field, as well as the driven displacement of the unipolar field . In this mode,  the  capacitive displacement meter is combined with a commercial ferroelectric tester. (b)  System operating in lock-in amplifier mode, allowing the  piezoelectric strain coefficients at a zero electric field to be accurately determined. }
\end{figure}

Figure \ref{Fig1} shows  a  block diagram of the system constructed for the strain  measurement. The system can operate in two modes. Figure \ref{Fig1}(a) shows the system running in normal mode, in which the  displacement meter is combined with a commercial ferroelectric tester (Toyo FE tester system,  FCE-3),  allowing  measurements of  the strain and dielectric D-E hysteresis loops simultaneously for a bipolar electric field, as well as a large unipolar-field-driven displacement. Figure \ref{Fig1}(b) show a schematic diagram of  the lock-in amplifier  mode,  which can be used to precisely  determine the  strain  piezoelectric coefficients $d_{33}$ at the zero field  from the linear relationship between the AC displacement and  voltage. It should be noted that the amplitude is  used to express the values of all AC signals in this study. 

As shown in Fig. \ref{Fig1},  the displacement is measured  by detecting the change in air capacitance $C$ that forms  between the capacitive displacement sensor and object to be measured. Here, we have
\begin{equation}\label{eq1}
    C=\varepsilon_0 S/d,
\end{equation}
where $\varepsilon_0$ is the vacuum permittivity, $S$ is the sensor area, and $d$ is the distance between the sensor and  object to be measured. When a voltage ($V$) with frequency $f$ is applied to the air capacitor,  the  applied voltage $V$ and the current $I$ flowing through the sensor have the following relationship

\begin{equation}\label{eq2}
    V=I/(2\pi f C)=d I /(2\pi f \varepsilon_0 S).
\end{equation}
If $I$ is maintained at a constant value, then $d$ is
proportional to the output  voltage $V$ of the capacitive sensor.  As shown in Fig. \ref{Fig1}, $d$ is changed simultaneously as the sample is deformed by the applied electric field. We  can therefore use this relationship to  measure the sample deformation driven by the applied electric field by  measuring the sensor  output voltage.  In our system, we use an Iwatsu ST-0403-50 capacitive sensor, which  can  probe the   displacement within a  range of $\pm$25  $\mu$m with  a resolution of 0.3 nm at 1 kHz,  a linearity of 0.1\%,   and  a frequency range from DC to 1 kHz.  The capacitive sensor is equipped with an Iwatsu displacement meter ST-3541, which can provide a voltage output for data record (a voltage-displacement conversion coefficient of 2.5 nm/mV).

We found that  sample mounting   is of critical importance for reliably  probing the sample displacement. Here, the
sample is held between two cylindrical  hemisphere electrodes with a diameter of 5 mm to provide a point contact between the  sample and   hemisphere  electrodes. We also  found  that a linear bush is extremely effective  in guiding the cylindrical shaft and providing a  smooth and  precise linear movement, allowing it to accurately measure  the sample displacement. As shown in Fig.  \ref{Fig1}, a soft spring with a spring constant of 0.2 N/mm is used to make  a reliable contact between the sample electrode and the hemisphere electrodes, and the compress force applied  to the sample  is estimated to be approximately 0.6 N, which is far lower than the commonly used value of approximately 10 N in a Berlincourt meter or a  $d_{33}$ meter \cite{Stewart}. This soft contact is extremely effective at preventing  the damage to the sample, particularly for a thin film.  A Z-stage is used to adjust the air capacitance distance  for setting  the zero  output when  the sample is not driven  by  an electric field. Essentially, if the sample can be held between the hemisphere electrodes, the measurement can be reliably performed. To confirm the reliable electrode contact directly by eye, the top electrode of the sample with a lateral size of larger than 1 mm is usually used for the measurements, whereas the bottom electrode is generally coated with a larger area for a convenient sample setup.

At present, the system can operate within a temperature range of   210  to 450 K. We use  silicon oil as a heating medium to achieve a homogeneous sample temperature. In addition, silicon oil  can  also prevent a high voltage discharge, allowing  the  application of 10,000 V to a bulk sample.  A  silicon rubber heater of 100 W is used to heat the sample, whereas  liquid N$_2$  is  used for cooling to low temperature. To prevent  water condensation below the  ice point, the sample chamber is  evacuated using  a vacuum pump. The  silicon oil applied and  the system can  both undergo a thermal variation within a  temperature range of 173 to 573 K. Thus, the measurement   temperature    can be expanded to a larger range.

 \section{System Operation }
 
\begin{figure}
\includegraphics[width=12cm]{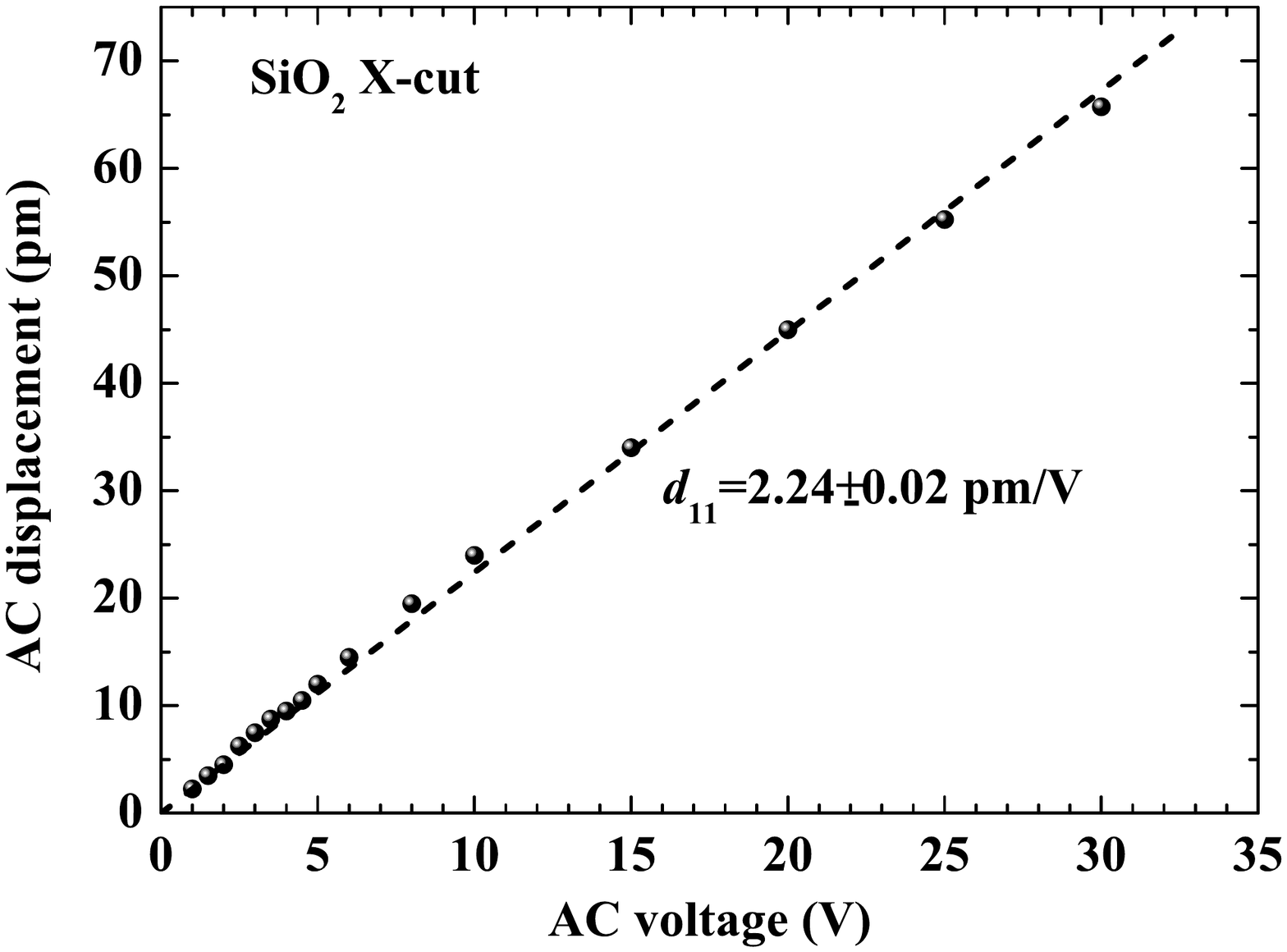}
\caption{\label{Fig2}  AC voltage driven displacement of the  single crystal X-cut quartz obtained by the lock-in amplifier technique at room temperature. The broken line shows a linear fitting for the calculation of  the  $d_{11}$ value of the crystal.  }
\end{figure}

First, we used   standard X-cut quartz to check the system operation in the lock-in amplifier mode. The X-cut quartz  was driven by a small AC voltage, and  the AC output of the  displacement meter was  fed into a lock-in amplifier (SIGNAL RECOVERY model 7265) to measure the AC  displacement.  As expected, a linear relationship  was observed between the displacement  and the applied voltage as shown in Fig. \ref{Fig2}. Using a linear fitting,   $d_{11}$ of   the X-cut quartz was evaluated to be 2.24 $\pm$ 0.02 pm/V, which is in extremely  good agreement with the reported values (2.27 $\pm$ 0.01 pm/V\cite{Bottom},  2.32 pm/V\cite{Pan}, 2.22 pm/V\cite{Graham},   2.27 $\pm$ 0.01 pm/V\cite{Uchino2}, and 2.314 pm/V\cite{Bhalla}). The results given in  Fig. \ref{Fig2}  also indicate that our system can  accurately  probe a 2-pm  displacement that is comparable to a $10^{-2} \AA$ resolution of a double-beam laser interferometer  \cite{Pan}, and far better than the X-ray diffraction technique.\cite{Bhalla} Because the  SIGNAL RECOVERY (model 7265) lock-in amplifier applied can provide a full-scale sensitivity of 2 nV, which corresponds to a displacement resolution of 5 fm, it is possible to improve the measurement resolution of our system.

\begin{figure}
\includegraphics[width=12cm]{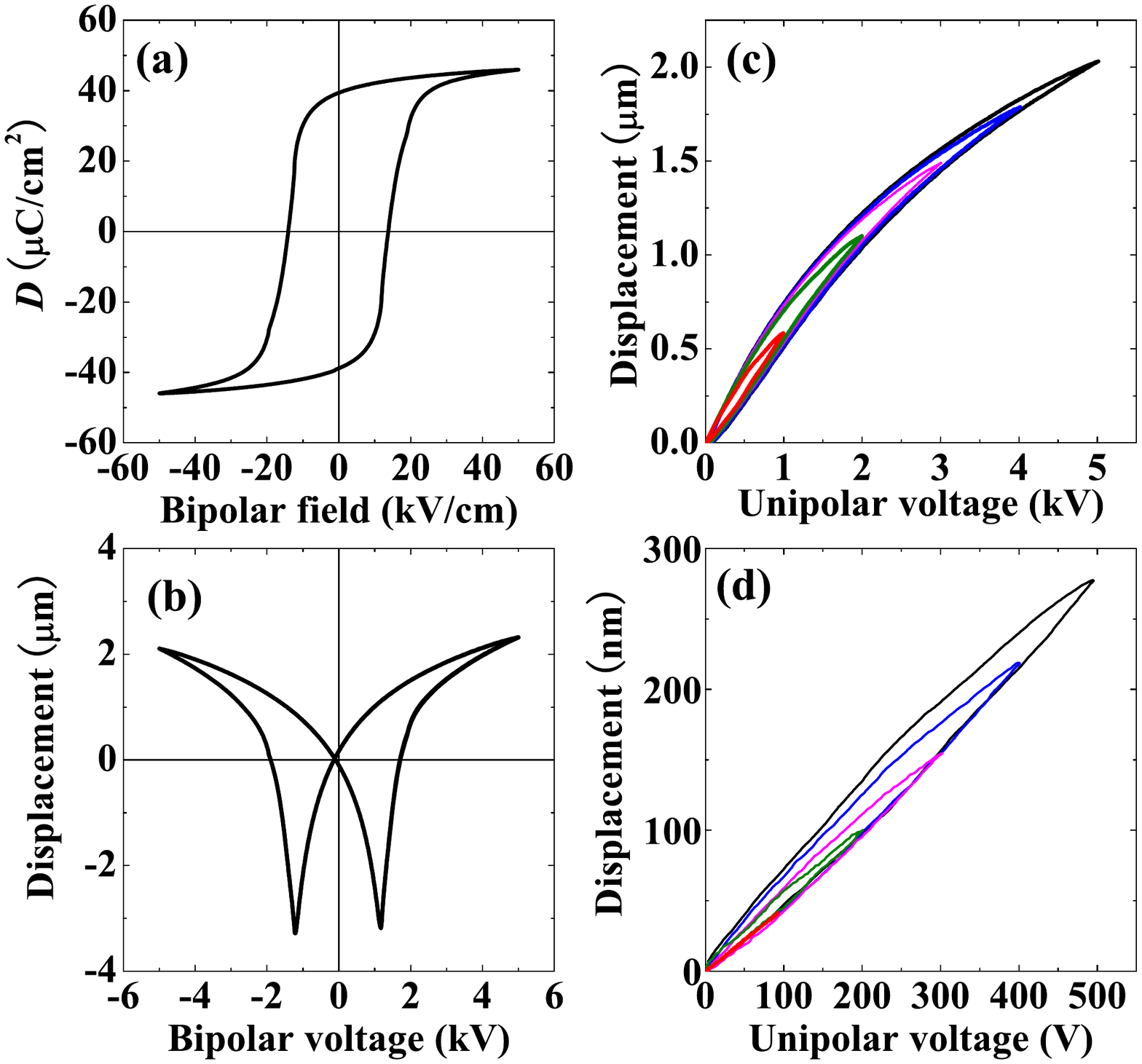}
\caption{\label{Fig3} (a) Dielectric  hysteresis loop of  industrial PZT
ceramics with a thickness of 1 mm under the application of a  bipolar electric
field at $T$ = 297 K. (b) Displacement hysteresis loop obtained
simultaneously with dielectric hysteresis loop. (c) Higher unipolar voltage driven displacements. (d) Lower unipolar voltage driven displacements.}
\end{figure}

We then used  commercial Pb(Zr$_{1-x}$Ti$_x$)TiO$_3$(PZT) ceramics with a diameter of 10 mm and a thickness of 1 mm to examine the system  operation in normal mode (without a lock-in amplifier) for measuring  the large  sample displacement. Figure  \ref{Fig3} shows the  simultaneously recorded dielectric and  displacement loops under the application of a bipolar field. From the  butterfly displacement loop, we can see that this PZT sample shows a  deformation of several micro-meters during polarization switch driven by the bipolar field. When  a unipolar  voltage of  5,000 V is applied to the samples,  it shows a displacement of approximately 2 $\mu$m,  as shown in Fig. \ref{Fig3} (c).   Both the non-linearity and hysteresis of the displacement are reduced by reducing the unipolar voltage. From Fig. \ref{Fig3}(d), a 45-nm  displacement can be read  at  a voltage of 100 V.  At this small voltage, the ratio between  displacement and  voltage is calculated as 450 pm/V, which is extremely close to the $d_{33}$  value of 441 pm/V obtained using  ZJ-6B Quasi-Static
Piezo $d_{33}$/$d_{31}$-meter made by the Institute of Acoustics, Chinese Academy of Sciences, which has a measurement accuracy of $\pm 1\%$.  Thus, we  can use the  ratio  between strain ($S$)  and  electric  field ($E$)  at a  low field (for example  $E$ = 1-2 kV/cm) to estimate  the piezoelectric coefficient $d_{33}$ of the sample. 

\section{Thin film measurements}
 
After examining  the system reliability when using  standard quartz and PZT ceramics, we then use it to study the displacement of a PZT thin film. The film used  has a composition of $x=0.5$ and a  thickness of 1 $\mu$m,   and was prepared through a chemical solution deposition technique on a silicon substrate with a Pt bottom  electrode.   A top Au electrode with a diameter of 1 mm was  then coated for the measurements. For easier measurements, the back side of silicon substrate without a PZT film was also coated by Au and connected to the Pt bottom electrode. 

\begin{figure}
\includegraphics[width=12cm]{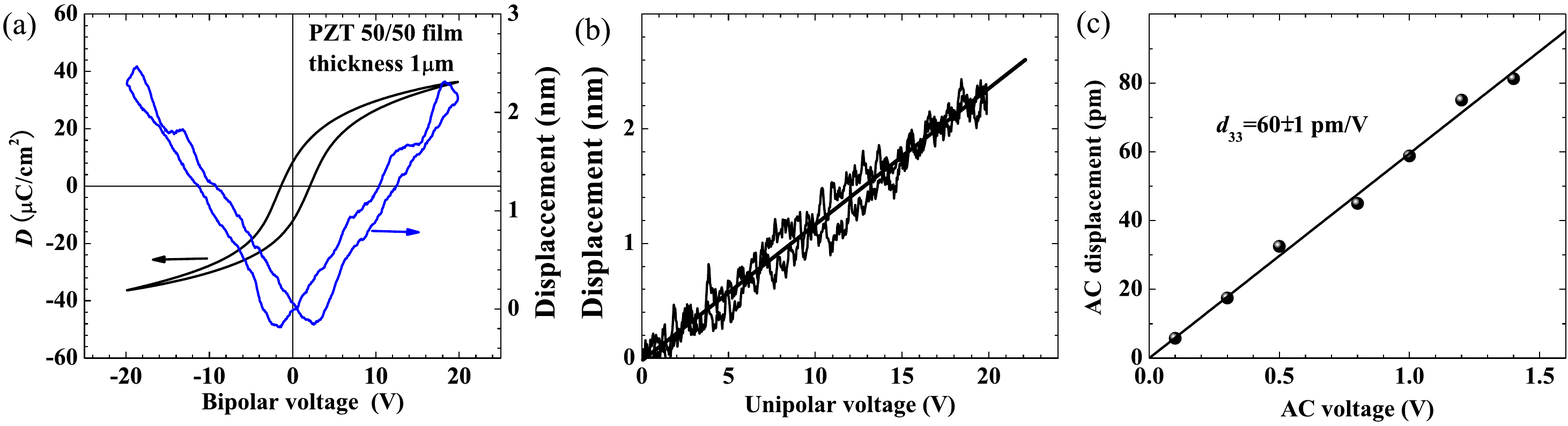}
\caption{\label{Fig4} (a) Dielectric and displacement  hysteresis loops of the PZT
film under the application of  bipolar voltage at room temperature. (b) The unipolar voltage driven displacement. The line is used as a guide.   (c)  The AC voltage driven displacements obtained using a  lock-in amplifier technique. The line shows a linear fitting for the calculation of  the $d_{33}$ value. }
\end{figure}

Figure \ref{Fig4}(a) shows the dielectric and displacement  hysteresis loops of the  PZT film. In a normal mode operation without a lock-in amplifier, it is difficult to measure a  nanometer-order displacement with a high resolution. To improve the resolution, we conducted 200 measurements and took the average, allowing us to obtain  a displacement hysteresis loop of a PZT film with an extremely good resolution, as shown in Fig. \ref{Fig4}(a).   This film shows a displacement of approximately 2.4  nm as driven by  a unipolar voltage of 20  V (corresponding to an electric field of   $E$ = 200 kV/cm),  as shown in Fig. \ref{Fig4}(b). The large  noise in the unipolar measurement  is  due to the smaller number of average measurements  (20).  We then  used  this unipolar measurement result to estimate the high field piezoelectric effect  of the PZT film, which shows   an  effective piezoelectric coefficient of  $ \sim$120 pm/V at  $E$ = 200 kV/cm.  In addition, it seems  that one can  directly apply a large field to drive  the film deformation for application without a  poling process.
 
We then measured the strain piezoelectric coefficient at a zero-field  after the  PZT film was poled by unipolar field measurements. An AC electric field lower than  the coercive fields $E_c$  was applied to the film to measure  the AC displacement using a lock-in amplifier mode. It should be noted that if the applied AC field is larger than  $E_c$, the polarization state of the poled  PZT film will be destroyed owing to the  polarization switching.    As  expected from the piezoelectric effects, a linear relationship was observed between the displacement and applied voltage in  Fig. \ref{Fig4}(c). A linear fitting  gives a  $d_{33}$ value of  $60 \pm 1$ pm/V for this PZT film. This value  is half  the value evaluated from the unipolar measurements at $E$ = 200 kv/cm.  This difference can be explained by the difference of the poled regions in the film between these two different  measurement techniques. The AC lock-in technique detects the piezoelectric response at the zero field of the remanent polarization after poling. By  contract, the unipolar field measurement probes the piezoelectric response at an extremely high field of 200 kV/cm,  which will result in larger poled regions in the film, and  thus a larger polarization than the remanent polarization after poling. This  is also  evident from the D-E  loop shown in Fig. \ref{Fig4}(a). In addition, the lower $d_{33}$ value of this PZT film as compared with the  ceramics samples is affected by many factors such as the  film quality, orientation,  and polarization relaxation.  When comparing the dielectric loop of the ceramics (Fig. \ref{Fig2}(a)) and that of  the film (Fig. \ref{Fig4}(a)), we can see that the remanent polarization $P_{\rm r} $ of the film ($P_{\rm r+} $ = +8 $\mu$C/cm$^2$, $P_{\rm r-} $ = -12 $\mu$C/cm$^2$)  is much lower than that of PZT ceramics ($P_{\rm r}$  of  $\sim$40 $\mu$C/cm$^2$). In addition, the polarization in a thin film  is unstable after removing the applied field.  However, the  $d_{33}$ value obtained here falls within the  reported range of the PZT  films.\cite{Setter,Taylor,Gerbe} The above results clearly indicate that the developed  system  is capable of probing the piezoelectric effects in a thin film.

\section{Temperature variation measurements}

Some   piezoelectric  applications may  involve environments where the
temperature is different from  room temperature. It is therefore necessary to characterize the behavior of the piezoelectrics  over a range of possible operating temperatures. There are some reports on the temperature dependence of the piezoelectric properties including the piezoelectric coefficients, the elastic compliance coefficients, and the  electromechanical coupling factors, which are commonly measured using a resonance method. \cite{Zhang,Yokosuka,ZhangS,Sabat} However, there is a lack of understanding of   the temperature variation of the electric-field induced strain in a piezoelectric, which is of critical importance for  actuator  applications. We thus developed  our system to be capable of measuring the piezoelectric strain within  a temperature range of  210 - 450 K, which is important for practical applications.

Here, we use the  PZT ceramics shown in the  above  measurements and BaTiO$_3$(BTO) ceramics to demonstrate  the temperature  operation of our system. The measurement is conducted during  the cooling process at a cooling rate of 1 K/min. For each temperature measurement, we first measure the dielectric and displacement loops under the application of  a bipolar field. We then apply  unipolar fields with  maximum bipolar field (50 kV/cm for PZT and 30 kV/cm for BTO), 10 kV/cm, and 2 kV/cm, respectively, to measure the displacement  dependence on the  field. 

\begin{figure}
\includegraphics[width=12cm]{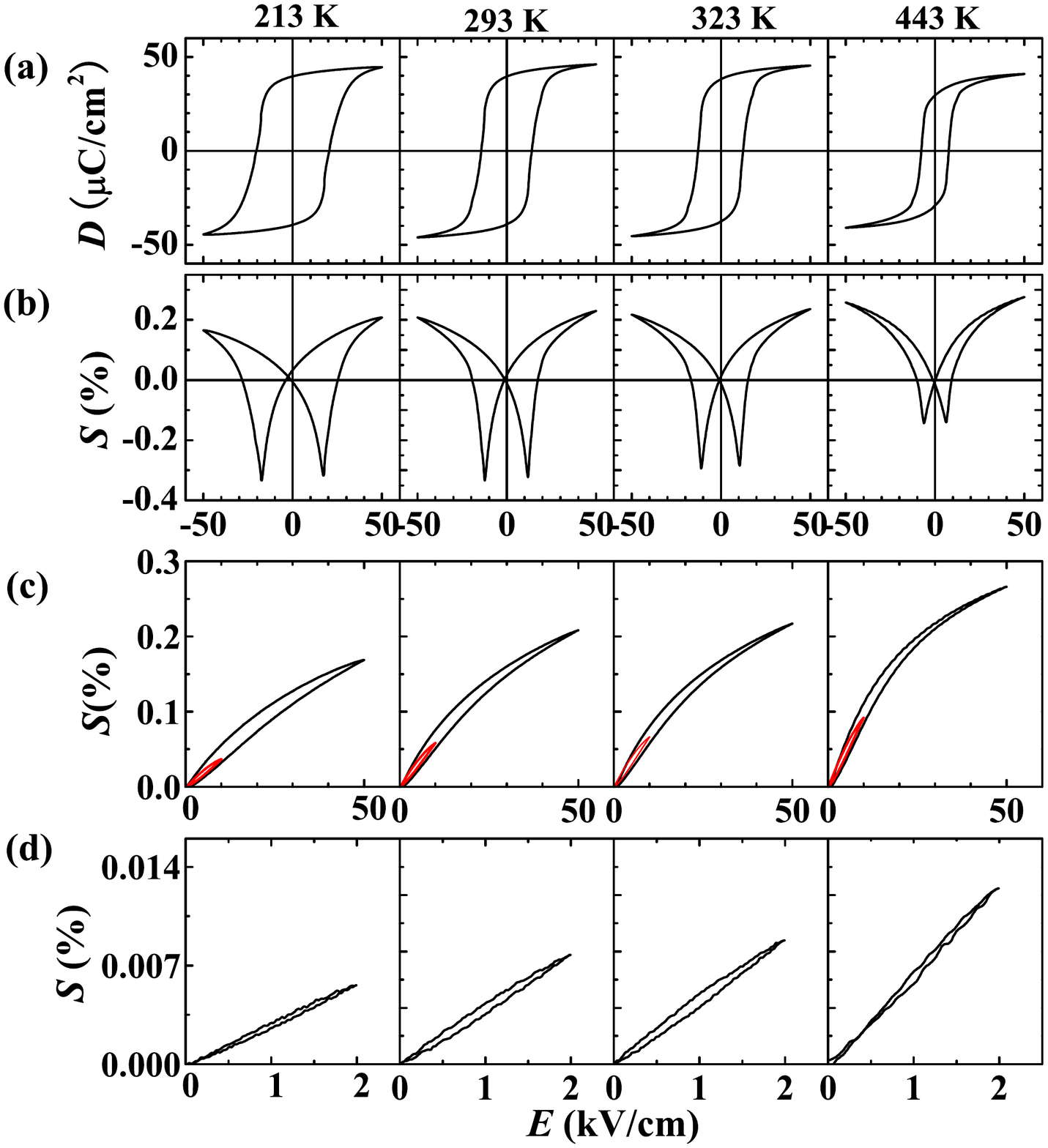}
\caption{\label{Fig5} Temperature dependence of (a) dielectric and (b) strain  hysteresis loops under the application of  bipolar electric  field, and  the unipolar field driven  strain response  at  (c) $E =$ 50 kV/cm and 10 kV/cm and  (d) 2 kV/cm for the PZT ceramics.}
\end{figure}

Figure \ref{Fig5} shows the temperature dependence of dielectric and strain  hysteresis loops under the application of a bipolar field, and the unipolar field driven strain of PZT ceramics. The used PZT ceramics  has a Curie point of 570 K (cooling)  determined  from the dielectric measurement.  It can been seen that the polarization increases by reducing the temperature. By contrast, the strain  response decreases upon cooling. Figures \ref{Fig5}  (c)  and (d) show the temperature variation of the  strain at a unipolar field of 50 kV/cm, 10 kV/cm, and 2 kV/cm. Here, we use the  ratio of the maximum strain ($S$) to  the maximum unipolar field ($E$) to evaluate the piezoelectric  strain response at different electric fields. The $S/E$ value can be considered  an effective piezoelectric coefficient in different fields. In particular,  the  $S/E$ value  at  $E$ = 2 kV/cm is extremely close to the $d_{33}$ value  obtained by a $d_{33}$-meter. 
 
\begin{figure}
\includegraphics[width=12cm]{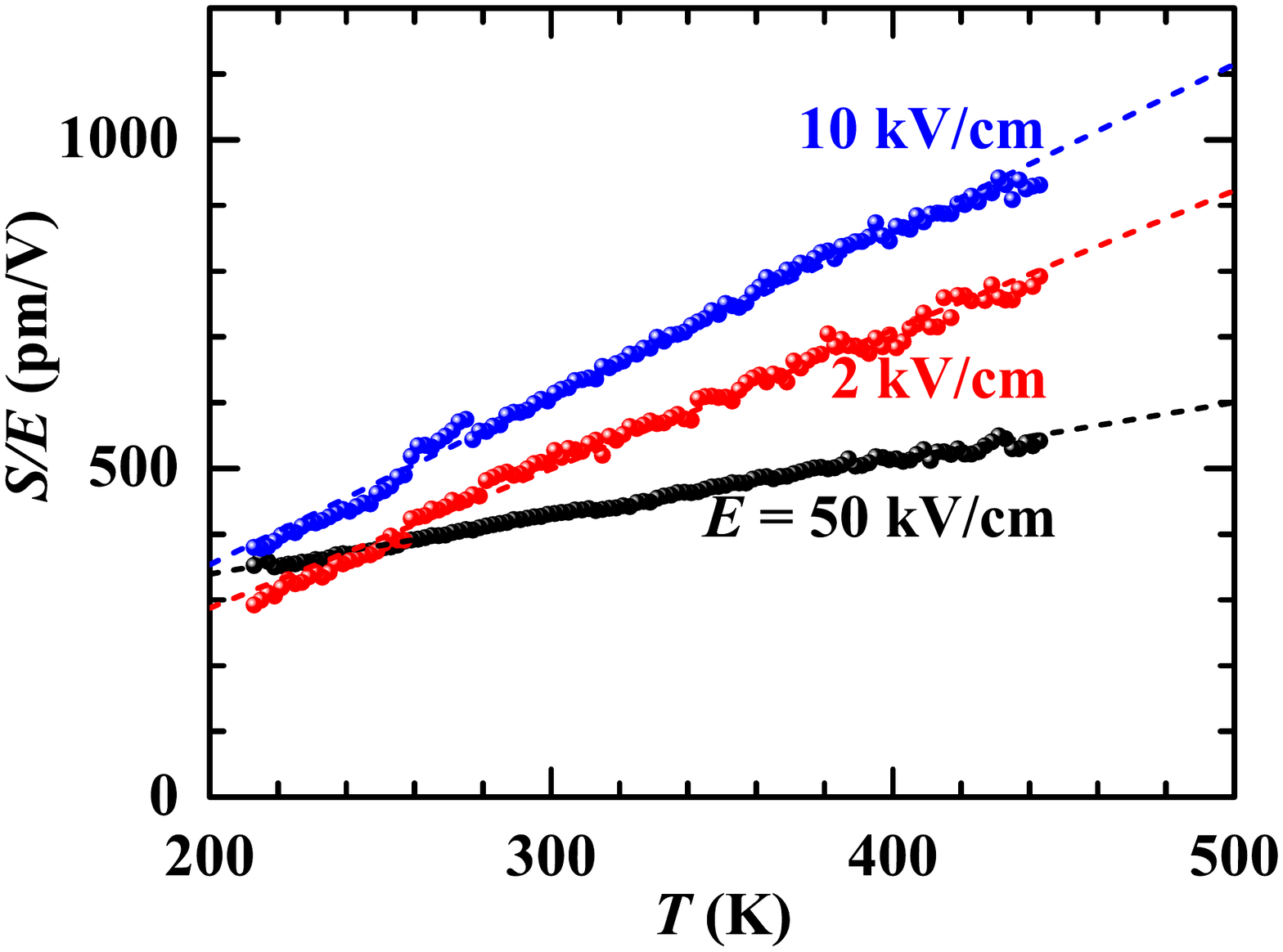}
\caption{\label{Fig6} Temperature dependence of the strain response at $E$ = 50 kV/cm, 10 kV/cm, and 2 kV/cm, respectively,  for the PZT ceramics. The broken lines show  a linear fitting for the calculation of  the temperature coefficient.}
\end{figure}

 Figure \ref{Fig6} shows the variation in $S/E$ with temperature for $E$ = 2 kV/cm, 10 kV/cm, and 50 kV/cm, respectively. Such a temperature dependence of PZT  ceramics  is consistent with that of  piezoelectric coefficients  $d_{33}$ of PZT ceramics obtained using  the resonance method.\cite{Zhang,Burianova} Among the three applied fields, the  $S/E$ value  at $E$ = 50 kV/cm is the  smallest. This can be attributed to the large saturation of domain motion under large electric field, as evident in  Fig. \ref{Fig5}(c). By contrast, the $S/E$ value  at  $E$ = 10 kV/cm is  the largest, which can be considered to result from the active domain motion in this field.  
 
 Herein, we use a linear fitting to estimate the temperature coefficient of  the $S/E$. The temperature coefficients are calculated to be 2.12 $\pm$ 0.02 pmV$^{-1}$K$^{-1}$,  2.54 $\pm$ 0.02 pmV$^{-1}$K$^{-1}$, and  0.870 $\pm$ 0.07 pmV$^{-1}$K$^{-1}$ for $E$ = 2 kV/cm, 10 kV/cm, and 50 kV/cm, respectively, within the measured temperature range. It should be noted that the noise in the temperature measurements is mainly due to a thermal shift  of the system.

\begin{figure}
\includegraphics[width=12cm]{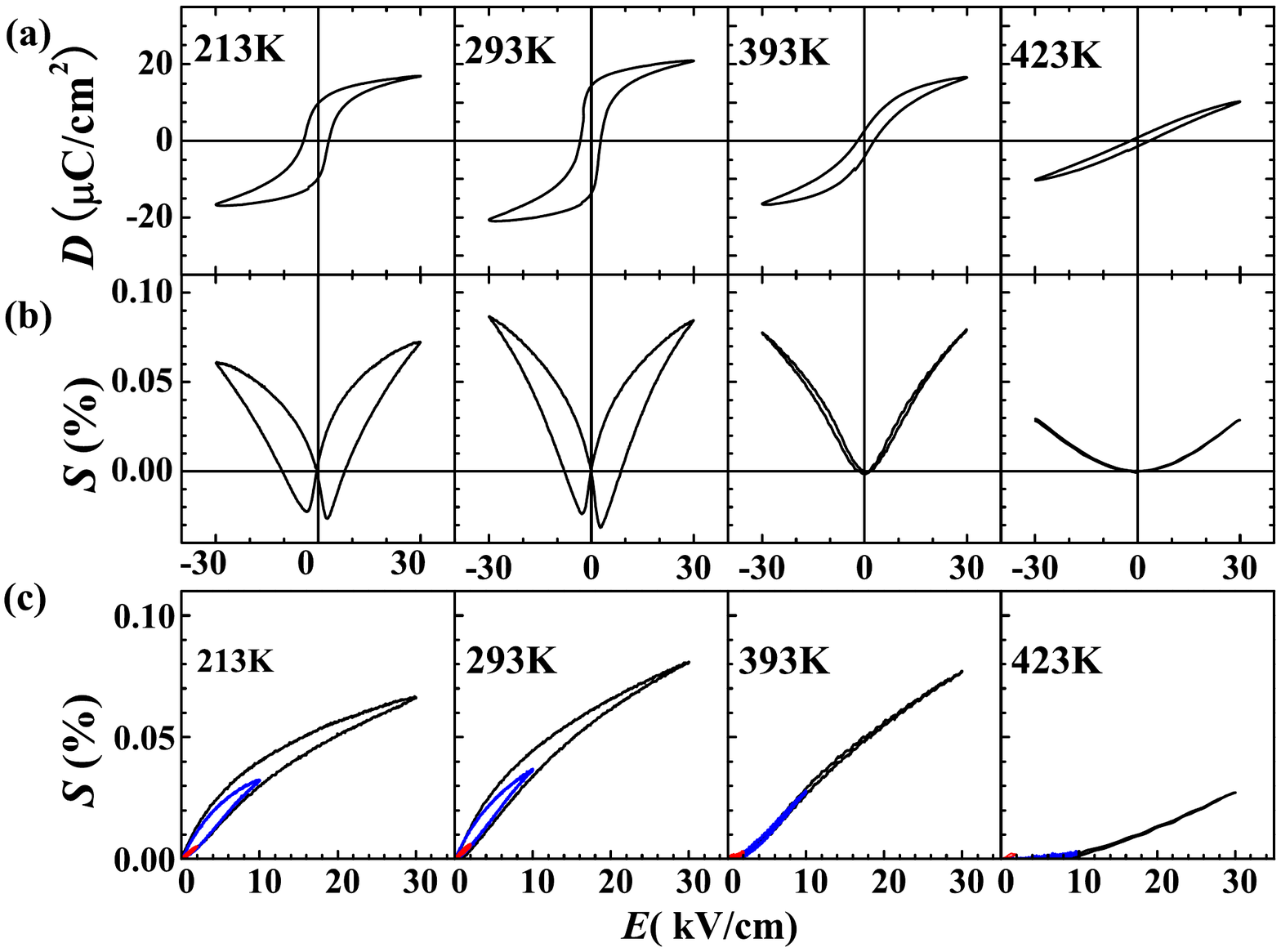}
\caption{\label{Fig7}  Temperature dependence of (a) dielectric and (b) strain  hysteresis loops under the application of  a bipolar electric  field, and (c) the unipolar field driven strain  at $E$ = 30  kV/cm, 10 kV/cm, and 2 kV/cm for BTO-Sn2.5\% ceramics. }
\end{figure}

We then demonstrate that the system can be used to investigate the influence of phase transition on the piezoelectric response by using BTO ceramics as an example.  To improve the insulation of BTO ceramics and  obtain a good dielectric loop at high temperature, 2.5 mol\% of Ti was substituted using Sn. Figure \ref{Fig7} shows the temperature dependence of dielectric and strain  hysteresis loops under the application of a  bipolar field, and the unipolar-field-driven  strain at  $E$ = 30 kV/cm, 10 kV/cm, and 2 kV/cm, respectively.  As shown in Fig. \ref{Fig7}, at $T$ = 423 K, BTO-Sn2.5\% is  in a  nearly paraelectric state although  a slightly nonlinear polarization appears at high electric  field (the slight open polarization loop may be due to the leakage current). Therefore, the strain response is due to the electrostrictive effect, which is evident from the relationship of $S\propto E^2$ in the strain response. When the sample is cooled to temperatures close to the Curie point  $T_{\rm c}$(= 380 K) such as $T$ = 393 K, the strain response is a mixture  of the piezoelectric and electrostrictive effects owing to the appearance of local polarization in the sample. After a phase transition ($T<T_{\rm c}$ ), BTO-Sn2.5\% is in a complete ferroelectric state,  showing  a typical butterfly strain loop for a bipolar field.

\begin{figure}
\includegraphics[width=12cm]{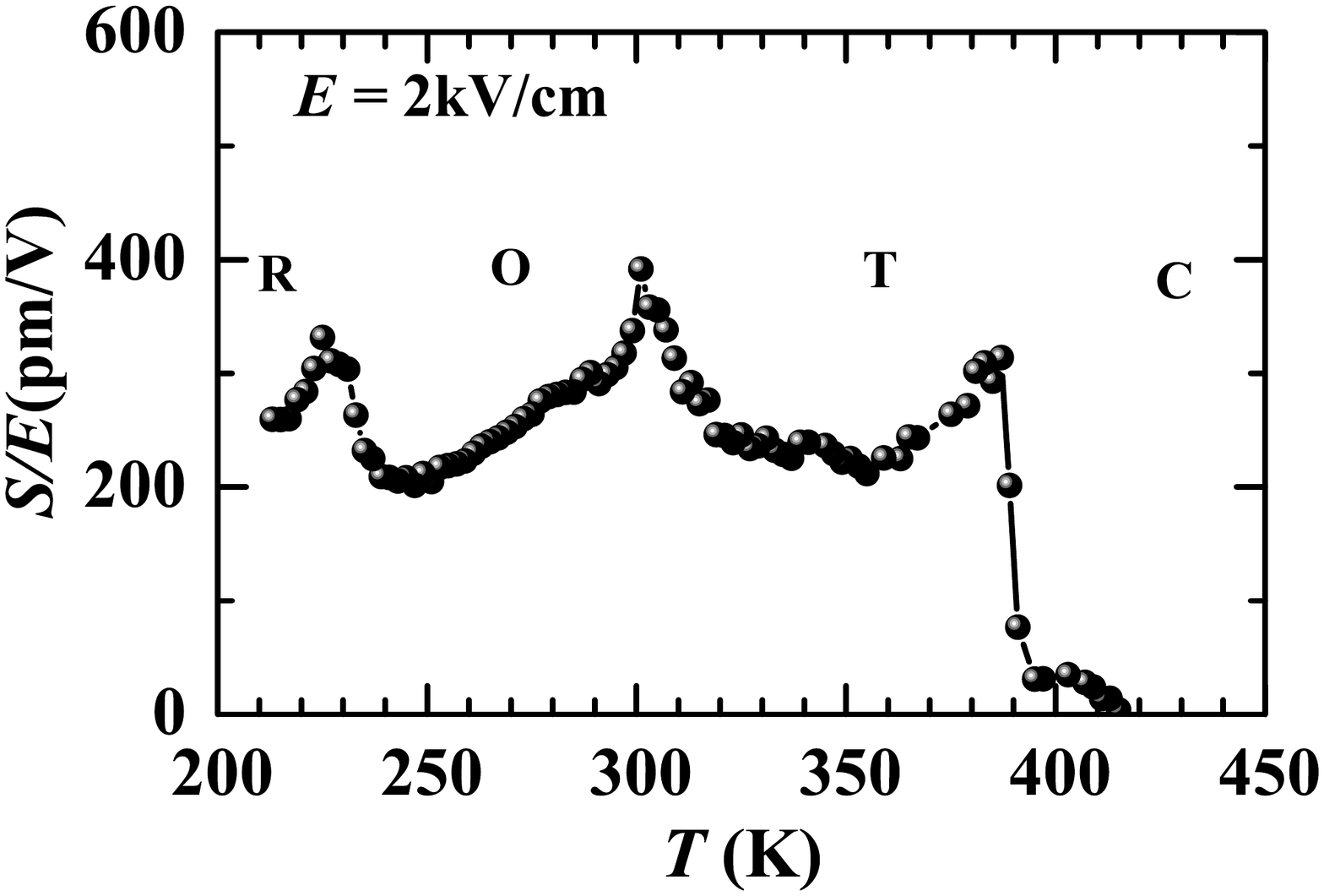}
\caption{\label{Fig8} Temperature dependence of the strain response at $E$ = 2kV/cm for  BTO-Sn2.5\% ceramics. }
\end{figure}

Figure \ref{Fig8} shows the temperature dependence of $S/E$ obtained at $E$ = 2 kV/cm. It can be seen that  strain  anomaly appears at  the successive  phase transitions in BTO-Sn2.5\%. At the phase  transition point, the strain response shows  the  maximum value. In addition, a larger strain response exists  at temperatures of approximately $T_{\rm c}$ within the cubic phase owing to the appearance of local polarization in the crystal. Similar phenomena have  also been  observed for the larger electric field of $E$ = 10 kV/cm and  $E$ =30 kV/cm.

In summary,  a  capacitive displacement  technique was developed, which allows  measuring the  electric-field-induced displacement in both bulk and  film piezoelectric materials.  The method  can accurately determine  an AC  displacement of as small as  2 pm  using a lock-in amplifier technique. It also allows measuring  the large displacement within $\pm$25  $\mu$m with a sub-nanometer resolution.  Moreover,  it can be used to investigate the strain variation within a temperature range of 210 - 450 K, allowing the temperature coefficient of the  piezoelectric strain  to be evaluated and the structure phase transition effects on the piezoelectric strain response to be studied. Experimental results on quartz, PZT ceramics and a thin film, and BaTiO$_3$ ceramics confirm the capabilities of this technique.


\begin{acknowledgments}
We thank Prof. H. Suzuki at Shizuoka University for providing the PZT thin film and Dr. R. Wang at AIST for  providing the commercial PZT ceramics sample used for the examination of our  system. 
\end{acknowledgments}

\textbf{DATA AVAILABILITY}

The data supporting  the findings of this study are available within this  article.


\clearpage
\begin{thebibliography}{50}%
\bibitem{Curie} J. Curie and  P.Curie, Bull. Soc. Fr.Mineral.
\textbf{3}, 90 (1880).
\bibitem{Jaffe} B. Jaffe, W. R. Cook Jr., and H. Jaffe, {\it Piezoelectric Ceramics}
(Academic, London, Great Britain, 1971).
\bibitem{Lines} M. E. Lines and A. M. Glass, {\it Principle and Application of Ferroelectrics and Related
materials} (Clarendon, Oxford, Great Britain,
1977).
\bibitem{Uchino} K. Uchino, {\it Piezoelectric Actuators and
Ultrasonic Motors} (Kluwer Academic Publishers, Norwell, MA 1997).


\bibitem{Muralt1995} P. Muralt, M. Kohli, T. Maeder, A. Kholkin, K. Brooks, N. Setter, and
R. Luthier, Sensors and Actuators A \textbf{48}, 157
(1995).

\bibitem{Muralt1997}P. Muralt, Integrated Ferroelectrics \textbf{17}, 297 (1997).
\bibitem{Damjanovic} D. Damjanovic, Rep. Prog. Phys.
\textbf{61}, 1267 (1998).
\bibitem{DeVoe} D. L. DeVoe, Sensors and Actuators A \textbf{88},
263 (2001).



\bibitem{ZhangS1} S. Zhang, F. Li, X. Jiang, J. Kim, J.  Luo, and  X. Geng,  Prog. Mater. Sci. \textbf{68}, 1 (2015).
\bibitem{Acosta} M. Acosta, N. Novak, V. Rojas, S. Patel, R. Vaish, J. Koruza, G. A. Rossetti, and J. Rodel,  Appl. Phys. Rev. \textbf{4}, 041305 (2017).
\bibitem{McKinstry} S. Trolier-McKinstry,. S. Zhang, A. J. Bell, and X. Tan, Annu. Rev. Mater. Res. \textbf{48}, 6 (2018).
\bibitem{Liu}H. Liu, J. Zhong, C. Lee, S.-W. Lee, and L. Lin, Appl. Phys. Rev. \textbf{5}, 041306 (2018).

\bibitem{FuAPL2007} D. Fu, M. Endo, H. Taniguchi, T. Taniyama, and M. Itoh, Appl. Phys. Lett.  \textbf{90}, 252907 (2007).
\bibitem{FuPRL2008} D. Fu, M. Itoh, S. Koshihara, T. Kosugi, and S. Tsuneyuki, Phys. Rev. Lett.  \textbf{100}, 227601 (2008).
\bibitem{FuAPL2008} D. Fu, M. Itoh, and S. Koshihara, Appl.  Phys. Lett. \textbf{93}, 012903 (2008).
\bibitem{Rodel} J. R\"{o}del, W. Jo , K. T. P. Seifert,
E. M. Anton, and T. Granzow, J. Am. Ceram. Soc. \textbf{92}, 1153 (2009).
\bibitem{FuJPC2011} D. Fu, M. Endo, H. Taniguchi, T. Taniyama, M. Itoh, and S. Koshihara, J. Phys.: Condens. Matter. \textbf{23},
075901 (2011).
\bibitem{FuAPL2011} D. Fu, T. Arioka, H. Taniguchi, T. Taniyama, and M. Itoh, Appl. Phys. Lett. \textbf{99}, 012904 (2011).
\bibitem{FuJPC2013} D. Fu, Y. Kamai, N. Sakamoto, N. Wakiya, H.
Suzuki and M. Itoh,  J. Phys.: Condens. Matter. \textbf{25}, 425901 (2013).

\bibitem{IEEE}  IEEE standard on piezoelectricity. ANSI/IEEE Stand, 176-1987.
\bibitem{Fialka} J. Fialka and P. Benes, IEEE Trans.  Instrum.  Meas.  \textbf{62}, 1047 (2013)
\bibitem{Berlincourt} D. A. Berlincourt, J. Appl. Phys. \textbf{30},1804 (1959).
\bibitem{Stewart} M. Stewart and  M. G.  Cain, {\it  Characterization of Ferroelectric Bulk Materials and Thin Films},  M. G.  Cain(Ed), (Springer, Dordrecht,  2014), 37-64.
\bibitem{Bottom} V. E. Bottom, J. Appl.  Phys. \textbf{41},  3941 (1970).
\bibitem{Pan} W. Y. Pan and L. E. Cross, Rev. Sci. Instrum. \textbf{60}, 2071 (1989).
\bibitem{Lefki} K. Lefki and G. J. M. Dormans,  J. Appl. Phys. \textbf{76}, 1764 (1994).
\bibitem{Kholkin} A. L. Kholkin, Ch. W\"{u}tchrich, D. V. Taylor, and N. Setter,  Rev. Sci. Instrum. \textbf{67}, 1935 (1996).
\bibitem{Kanno}I. Kanno, S. Fujii, T. Kamada, and R. Takayama, Appl. Phys. Lett. \textbf{70}, 1378 (1997).
\bibitem{Christman} J. A. Christman, R. R. Woolcott, Jr., A. I. Kingon, and R. J. Nemanich, Appl. Phys. Lett. \textbf{73}, 3851 (1998).
\bibitem{Xu} F. Xu, F. Chu, and S. Trolier-McKinstry, J. Appl. Phys. \textbf{86}, 588 (1999).
\bibitem{Durkan} C. Durkan, D. P. Chu, P. Migliorato, and M. E. Welland, Appl. Phys. Lett. \textbf{76}, 366 (2000).
\bibitem{Kuffer}  O. Kuffer, I.Maggio-Aprile, J.-M. Triscone, O. Fischer, and Ch. Renner,  Appl. Phys. Lett. \textbf{77},  1701 (2000).
\bibitem{FuJJAP} D. Fu, K. Ishikawa, M. Minakata, and H. Suzuki, Jpn. J. Appl. Phys. \textbf{40}, 5683 (2001).
\bibitem{Park} G.-T. Park, J.-J. Choi, J. Ryu, H. Fan, and H.-E. Kim, Appl. Phys. Lett. \textbf{80}, 4606 (2002).



\bibitem{Graham} G. M. Graham and F. N. D. D. Pereira, J. Appl. Phys. \textbf{42}, 3011  (1971).
\bibitem{Uchino2} K. Uchino and  L. E. Cross, Ferroelectrics \textbf{35}, 3941 (1980).
\bibitem{Bhalla} A. S. Bhakka, D. N. Bose, E. W. White, and L. E. Cross,  phys.  stat. sol. (a) \textbf{7}, 335 (1971).
\bibitem{Setter} N. Setter, D. Damjanovic, and  L. Eng, J. Appl. Phys.  \textbf{100}, 051606 (2006).
\bibitem{Taylor} David V. Taylor and D. Damjanovica, Appl.  Phys. Lett. \textbf{76},  1615 (2000).
\bibitem{Gerbe} P. Gerber, A. Roelofs, O. Lohse, C. Kugeler, S. Tiedke, U. Bottger, and R. Waser,  Rev.
Sci. Instrum. \textbf{74}, 2613 (2003).
\bibitem{Zhang} Q. M. Zhang,  H. Wang,N.  Kim, and  L. E. Cross, J. Appl. Phys. \textbf{75}, 454 (1994). 
\bibitem{Yokosuka} M. Yokosuka, T. Ochiai, and M. Marutake, Jpn. J. Appl. Phys.  \textbf{30}, 2228 (1991). 
\bibitem{ZhangS} S. Zhang, X. Dong, and S. Kojima, Jpn. J. Appl. Phys.  \textbf{36},  2994 (1997).
\bibitem{Sabat} R. G. Sabat, B. K. Mukherjee, W. Ren, and  G. Yang, J. Appl. Phys. \textbf{101}, 064111 (2007).
\bibitem{Burianova} L. Burianova, P. Hana, M. Pustka, M. Prokopova, and J. Nosek, J.  Euro.  Ceram. Soc.  \textbf{25}, 2405 (2005).








\end{thebibliography}

%

\clearpage

\clearpage

\clearpage

\clearpage

\clearpage

\clearpage

\clearpage

\clearpage



\end{document}